\documentclass[]{emulateapj}
\usepackage{CJK}
\usepackage{amsmath}
\usepackage{natbib}
\usepackage{graphicx}
\usepackage{epstopdf}
\usepackage{txfonts}
\usepackage{subfigure}

\newcommand{\beq}[1]{\begin{equation}\label{#1}}
\newcommand{\eeq}{\end{equation}}

\shorttitle{How Empty Are Gaps Opened by Planets?}
\shortauthors{Fung, Shi, \& Chiang}

\begin{document}
\begin{CJK*}{UTF8}{bsmi}
\title{HOW EMPTY ARE DISK GAPS OPENED BY GIANT PLANETS?}

\author{Jeffrey Fung (馮澤之)\altaffilmark{1}, Ji-Ming Shi (史集明)\altaffilmark{2}, and Eugene Chiang (蔣詒曾)\altaffilmark{2,3}}

\altaffiltext{1}{Department of Astronomy and Astrophysics, University of Toronto, 
50 St.~George Street,
    Toronto, Ontario, Canada M5S 3H4}
\altaffiltext{2}{Department of Astronomy, UC Berkeley, Hearst Field Annex B-20,
    Berkeley, CA 94720-3411}
\altaffiltext{3}{Department of Earth and Planetary Science, UC Berkeley, 307 McCone Hall,
    Berkeley, CA 94720-4767}

\email{Electronic address: fung@astro.utoronto.ca}

\begin{abstract}
  Gap clearing by giant planets has been proposed to explain the
  optically thin cavities observed in many protoplanetary disks. How
  much material remains in the gap determines not only how detectable
  young planets are in their birth environments, but also how strong
  co-rotation torques are, which impacts how planets can survive fast
  orbital migration. We determine numerically how the average surface
  density inside the gap, $\Sigma_{\rm gap}$, depends on
  planet-to-star mass ratio $q$, Shakura-Sunyaev viscosity parameter
  $\alpha$, and disk height-to-radius aspect ratio $h/r$. Our results
  are derived from our new GPU-accelerated Lagrangian hydrodynamical
  code \texttt{PEnGUIn}, and are verified by independent simulations
  with \texttt{ZEUS90}.  For Jupiter-like planets, we find
  $\Sigma_{\rm gap}\propto q^{-2.2}\alpha^{1.4}(h/r)^{6.6}$, and for
  near brown dwarf masses, $\Sigma_{\rm gap}\propto
  q^{-1}\alpha^{1.3}(h/r)^{6.1}$. Surface density contrasts inside and
  outside gaps can be as large as $10^4$, even when the planet does not accrete.
  We derive a simple analytic scaling, $\Sigma_{\rm gap}
  \propto q^{-2} \alpha^1 (h/r)^5$, that compares reasonably well to
  empirical results, especially at low Neptune-like masses, and use
  discrepancies to highlight areas for progress.
\end{abstract}


\section{INTRODUCTION} \label{intro}

Observational studies of giant planet formation will begin in earnest
once we detect planets still embedded in their natal gas disks.
Directly imaging young gas giants is made easier by their ability to
clear material away from their orbits. Planetary (Lindblad) torques
open gaps while viscous torques fill them back in \citep{goldreich80}:
a balance between these torques sets the equilibrium surface density
near the planet.
There are hints of gap clearing by planets in 
so-called ``transitional'' disks having optically thin cavities
\citep[e.g.,][]{kraus12,Debes13,Quanz13}.
Transition disk holes are surprisingly large; one gap opened
  by a single planet would be too narrow to explain the observed
  cavity sizes that range up to $\sim$100 AU, and so a given system
might have to contain multiple super-Jovian planets to clear a wide
enough swath \citep[e.g.,][]{Zhu11,dodson11}.  Even so, the
holes are so optically thin that planets alone seem incapable of
torquing material strongly enough to compete with viscous diffusion
--- at least for typically assumed parameters --- and appeals are made
to opacity reductions through grain growth or dust filtration at the
outer gap edge \citep{Zhu12,Dong12}.  Part of the motivation of our
study is to expand the parameter space explored and see how empty a
gap can be cleared.

Determining gap surface densities and corresponding optical depths is
relevant not only for observations but also for theory: material that
co-rotates with the planet (executing quasi-horseshoe orbits) can
backreact gravitationally on the planet and influence its dynamical
evolution.  The co-rotation torque takes its place among the litany of
resonant planet-disk interactions that can alter orbital
eccentricities and semimajor axes (for a review, see
\citealt{kley12}).  The delicate balance of forces within the gap,
including the thermodynamic behavior of matter there, may determine
how low-mass and giant planets survive the threats of Types I and II
orbital migration (\citealt{ward97}; see also section 2.2 of
\citealt{kley12} and references therein).

Despite its importance, $\Sigma_{\rm gap}$ --- the surface density
averaged over the bottom of the gap --- remains poorly
understood. Notwithstanding the huge number of simulations of
planet-disk interactions published in the past two
decades,\footnote{Much of the literature is marked by a peculiar
  insistence on plotting surface density $\Sigma$ on a linear
  scale. The practice is unhelpful since surface density contrasts in
  and out of gaps can span orders of magnitude --- indeed they must if
  they are to reproduce the enormous optical depth contrasts inferred
  from observations of transition disks (e.g., \citealt{Dong12}).}  a systematic parameter
study has yet to be performed that determines $\Sigma_{\rm gap}$ as a
function of planet-to-star mass ratio $q \equiv M_p/M_\ast$;
Shakura-Sunyaev viscosity parameter $\alpha$; and disk
height-to-radius aspect ratio (equivalently, disk temperature)
$h/r$. \citet{crida06} examined how these parameters influence gap
shape and width, but not gap depth --- i.e., they studied the onset of
the gap, but not the bottom of the gap. The numerics can be
challenging. Measuring $\Sigma_{\rm gap}$ accurately requires global
simulations that 
(i) resolve the disk well in at least azimuth $\phi$ and radius $r$;
(ii) resolve large density contrasts;
(iii) converge with time to a steady state;
(iv) model how the planet accretes from the ambient flow; and 
(v) possess well-separated radial boundaries that maintain a steady mass accretion
rate across the entire domain, i.e., the simulation presumably should enforce
$\dot{M}(r) =$ constant $\neq 0$, as befits real accretion disks. 
Feature (v) is captured by only a minority of studies, and feature (iv) 
can only be mocked up in a parameterized way
\citep[e.g.,][]{lubow06,Zhu11}.

This paper aims to provide an empirical relation for $\Sigma_{\rm gap}
(q,\alpha, h/r)$ for a single non-accreting giant planet on a fixed
circular orbit embedded in a 2D, locally isothermal, steadily
accreting disk.  
We restrict our study to disk gas only, and ignore how dust 
and gas flows might differ.
We utilize two independent codes: \texttt{ZEUS3D}
\citep{stone92a}, and \texttt{PEnGUIn}, a new Lagrangian PPM
(piecewise parabolic method)-based code that we have implemented on
multiple GPUs (graphics processing units). To the extent possible,
results from one code will be validated against the other.

\subsection{An Analytic Scaling Relation} \label{sec:scaling}

Although our study is primarily numerical, we derive here an
approximate analytic relation for $\Sigma_{\rm gap}$ that we will use
to put our numerical results in context. We admit at the
  outset that our derivation can hardly be called such, as our
  reasoning will ignore many details and make assumptions not
  carefully justified. But as the rest of our paper will show, the
  simple relation we now present will yield results surprisingly close
  to those of detailed numerical simulations.

First examine the outer disk, exterior to the planet's orbit.
The outer Lindblad torque exerted by the planet transmits
angular momentum outward at a rate:
\begin{equation} \label{eqn:lindblad}
T_{\rm L} \sim q^2 \left( \frac{r}{h} \right)^3 \Sigma_{\rm gap} \,\Omega^2 r^4
\end{equation}
where $\Omega$ is the disk angular frequency \citep{goldreich80}.
This is the total torque from linear perturbations to one side of the
planet, integrated over all resonances up to the torque cut-off at
azimuthal wavenumber $m \sim r/h$ (see, e.g., equation 2 of
\citealt{crida06}, and references therein). Note that we have used
$\Sigma_{\rm gap}$, the surface density averaged over the bottom of
the gap, in our evaluation of the integrated Lindblad torque.
Our justification for this choice is that the integrated
torque is dominated by resonances at the torque cut-off, i.e., at
distances $\sim$$h$ from the planet (see figures 2 and 3
of \citealt{goldreich80}), and bottoms of gaps opened by giant
planets in gas disks typically extend this far (see, e.g.,
the simulated gaps of \citealt{crida06}, or our Figure \ref{fig:res}).
One caveat is that the linear theory from which equation
  (\ref{eqn:lindblad}) originates is formally valid only for low-mass
  planets for which $q \lesssim (h/r)^{3}$; the highest-mass planets
  we simulate will violate this condition, and indeed for such
  super-Jovian objects our numerical simulations will reveal deviations from the
  simple-minded scaling law we derive in this section.

In steady state, the outward transmission of angular momentum by the
(outer) Lindblad torque must be balanced by the angular momentum
transmitted inward
by the viscous torque:\footnote{There is also a so-called
    ``pressure'' torque of comparable magnitude to the Lindblad and
    viscous torques \citep{crida06}, but we neglect this third torque
    for our order-of-magnitude derivation.}
\begin{equation} \label{eqn:viscous}
T_{\rm v} \sim \Sigma_0 \,\nu \,\Omega \,r^2
\end{equation}
where $\nu = \alpha (h/r)^2 \Omega r^2$ is the kinematic viscosity
\citep[e.g.,][]{franketal02}.  
Here we have used $\Sigma_0$, 
the surface density at the gap periphery --- or more
  conveniently, the surface density at the planet's location if the
  planet were massless --- to evaluate $T_{\rm v}$. The viscous torque
  depends on the gradient of $\Sigma$, and this gradient is larger
  outside the flat-bottomed gap than inside it (see, e.g., the gap
  profile shown in Figure \ref{fig:res}; by ``gap periphery'' we mean
  a location like $r \approx 1.4$, where the gradient might reasonably
  be approximated as $\Sigma_0/r$, which is what equation
  \ref{eqn:viscous} essentially assumes).  Conscripting $\Sigma_0$ in
  this way is a gross simplification, but alternatives would require
  that we actually compute the precise shape of the gap, which is what
  we are trying to avoid with our order-of-magnitude derivation.

Usually one thinks of viscous torques as transmitting angular momentum
outward, but in the gap edge of the outer disk, the direction of viscous
transport is inward because the surface density there has a sharp and
positive gradient ($d\Sigma/dr > 0$).

Setting $T_{\rm L} = T_{\rm v}$ yields
\begin{equation} \label{eqn:scaling}
\frac{\Sigma_{\rm gap}}{\Sigma_0} 
\sim \frac{\alpha (h/r)^5}{q^2} \,.
\end{equation}
Exactly the same scaling relation applies to the inner disk, interior
to the planet's orbit. The signs in the inner disk are reversed from
those in the outer disk: the inner Lindblad
torque transmits angular momentum inward, while the local viscous torque
transmits angular momentum outward.

As we were completing our numerical tests of equation
(\ref{eqn:scaling}) and preparing our manuscript for publication, we
became aware of the study by \citet{duffell13} who found the same
scaling relation on purely empirical grounds (although these authors
did not explicitly vary the Mach number $r/h$). \citet{duffell13}
concentrated on the low-mass $q \lesssim 10^{-4}$ regime. Our study
complements theirs by studying the high-mass $q \gtrsim 10^{-4}$
regime; we will see to what extent equation (\ref{eqn:scaling}) also
holds true for giant planets. See also our \S\ref{sec:analytic} where we
discuss to what extent the short derivation given in this subsection
captures the whole story.

\subsection{Plan of This Paper} \label{sec:plan} Section
\ref{sec:method} contains our numerical methods and simulation
parameters. Section 3 presents our results for $\Sigma_{\rm
  gap}(q,\alpha,h/r)$. Section 4 concludes and charts directions for
future work.

\section{NUMERICAL METHODS}\label{sec:method}
We numerically simulate a planet on a fixed circular orbit embedded in
a co-planar, viscously accreting disk. Our two independent codes,
\texttt{PEnGUIn} and \texttt{ZEUS3D}, solve the usual mass and
momentum equations in two dimensions; we give here the equations in the
inertial, barycentric frame (although neither code actually works in this frame;
see \S\ref{sec:penguin} and \S\ref{sec:zeus} for the technical
details):

\begin{equation}
\label{eqn:cont_eqn}
\frac{D\Sigma}{Dt} + \Sigma\left(\nabla\cdot\mathbf{v}\right) = 0 \,,
\end{equation}

\begin{equation}
\label{eqn:moment_eqn}
\frac{D\mathbf{v}}{Dt} = -\frac{1}{\Sigma}\nabla P + \frac{1}{\Sigma}\nabla\cdot\mathbf{T}  - \nabla \Phi \,,
\end{equation}

\noindent
where $D/Dt$ is the Lagrangian derivative, $\Sigma$ is the surface
density, $P$ is the vertically integrated pressure, $\mathbf{v}$ is
the velocity, $\mathbf{T}$ is the Newtonian viscous stress tensor, and
$\Phi$ is the gravitational potential of the central star and planet
(but not the disk). In polar coordinates,
$\mathbf{v}=(v_{r},\Omega r)$; in component form,
equation~(\ref{eqn:moment_eqn}) reads:

\begin{align}
\nonumber
\frac{D v_{r}}{Dt} =& -\frac{1}{\Sigma}\frac{\partial P}{\partial r} + \frac{2}{\Sigma r}\frac{\partial}{\partial r}\left(\nu \Sigma r \frac{\partial v_{r}}{\partial r}\right) \\
&+ \frac{1}{\Sigma r}\frac{\partial}{\partial \phi}\left[\nu \Sigma \left(r\frac{\partial \Omega}{\partial r} + \frac{1}{r}\frac{\partial v_{r}}{\partial \phi} \right)\right] - \frac{\partial \Phi}{\partial r} \,,\\
\nonumber
\frac{D (r\Omega)}{Dt} =& -\frac{1}{\Sigma r}\frac{\partial P}{\partial \phi} + \frac{2}{\Sigma r}\frac{\partial}{\partial \phi}\left(\nu \Sigma \frac{\partial \Omega}{\partial \phi}\right) \\
&+ \frac{1}{\Sigma r^2}\frac{\partial}{\partial r}\left[\nu \Sigma r^2 \left(r\frac{\partial \Omega}{\partial r} + \frac{1}{r}\frac{\partial v_{r}}{\partial \phi} \right)\right] - \frac{1}{r}\frac{\partial \Phi}{\partial \phi} \,.
\end{align}

\noindent
Here $\nu = \alpha c_{\rm s} h$ is the kinematic viscosity following
\citet{alpha}, with $c_{\rm s}$ equal
to the sound speed.  We complete the equation set with a locally
isothermal equation of state $P = \Sigma c_{\rm s}^2$, with $c_{\rm s}
\propto r^{-1/2}$ so that the disk aspect ratio $h/r = $ constant.

In the center-of-mass frame,
\begin{align}
\label{eqn:potential}
\nonumber
\Phi =& -\frac{GM_*}{\sqrt{r^2 + r_{\rm 1}^2 + 2 r r_{\rm 1}\cos(\phi-\phi_{\rm p})}} \\
&-\frac{GM_{\rm p}}{\sqrt{r^2 + r_{\rm 2}^2 - 2 r r_{\rm 2}\cos(\phi-\phi_{\rm p})+ r_{\rm s}^2}},
\end{align}
where $M_\ast$ and $M_{\rm p} = q M_\ast$ are the masses of the star and the planet,
respectively;
 $r_{1} = q r_{\rm p}/(1+q)$ and $r_{2}=r_{\rm p}/(1+q)$ are their radial positions, with $r_{\rm p}$ the total (fixed) separation; 
$\phi_{\rm p}-\pi$ and $\phi_{\rm p}$ are their angular positions; and $r_{\rm s}$ is the smoothing (a.k.a.~softening)
length of the planet's potential. 
We set $G(M_\ast + M_{\rm p})=1$
and $r_{\rm p}=1$ so that the planet's orbital
frequency $\Omega_{\rm p}=1$ and period $P_{\rm p}=2\pi$.

\subsection{\texttt{PEnGUIn}: Code Description}
\label{sec:penguin}
\texttt{PEnGUIn} ({\bf P}iecewise Parabolic Hydro-code {\bf En}hanced
with {\bf G}raphics Processing {\bf U}nit {\bf I}mplementatio{\bf n})
is a Lagrangian, dimensionally-split, shock-capturing
hydrodynamics code. We defer a detailed description to a future paper,
and here just mention a few salient points.  The code is written in
CUDA-c and runs on multiple GPUs (graphics processing units) for
accelerated performance. It uses the piecewise parabolic method (PPM;
\citealt{PPM}), and its main solver is modelled after VH-1 \citep{VH1}
with two main differences: (i) \texttt{PEnGUIn} explicitly conserves
angular momentum by replacing, as one of the quantities
a fluid element carries in the Lagrangian frame, the angular speed
with the specific angular momentum, and (ii) \texttt{PEnGUIn}
uses a non-iterative Riemann solver for isothermal flows; see
\citet{Balsara94}, and note that their Lagrangian re-map formulation
applies not only to strictly isothermal flows, but also to locally
isothermal flows under the two-shock approximation by allowing for
different sound speeds for left- and right-moving
waves. \texttt{PEnGUIn} also contains a module that computes the
divergence of the stress tensor through piecewise parabolic
interpolations.

Technically, \texttt{PEnGUIn}'s reference frame is a barycentric frame
that rotates at $\Omega_{\rm p}$; thus the planet's position is fixed
in time, but the Coriolis force is not computed as an explicit source
term. Rather, it is absorbed into the conservative form of the angular
momentum equation \citep{Kley98}.

GPU-acceleration is optimal for programs that are massively parallel;
\texttt{PEnGUIn} achieves parallelization through memory management
and domain splitting. The hardware used for this paper are three
GTX-Titan graphics cards all connected to a single node. Running in
double precision on all three cards simultaneously, \texttt{PEnGUIn}
takes 12 seconds to run per planetary orbit for $(q,\alpha,h/r) =
(10^{-3},10^{-2},0.05)$.

\subsection{\texttt{ZEUS90}: Code Description}
\label{sec:zeus}
For comparison with \texttt{PEnGUIn}, we also carried out simulations
with \texttt{ZEUS90}: a modern version of \texttt{ZEUS3D}
\citep{stone92a,HS95} written in FORTRAN~90. It is a
three-dimensional, operator-split, time-explicit, Eulerian
finite-differencing magnetohydrodynamics code, widely used to simulate
a variety of systems, including magnetorotationally-unstable
circumbinary disks \citep{Shi2012} and warped disks
\citep{Sorathia2013}.  For our application, we suppress the vertical
dimension and magnetic fields; implement the Navier-Stokes viscosity
module; and add a planetary potential having a specified time
dependence.  The von Neumann-Richtmyer artificial viscosity,
commonly used to capture shock waves, is switched off in the
presence of an explicit viscosity.  The reference frame for
\texttt{ZEUS90} is a non-rotating frame centered on the star, and so
ordinarily there is an extra term in $\Phi$ due to the indirect
potential: $GM_{\rm p} r \cos(\phi-\phi_{\rm p})/r_{\rm p}^{2}$.
However, we find in practice that the indirect term results in a
``wobbling'' of the disk that is difficult to reconcile with our fixed
boundary conditions on fixed circles (see equations
\ref{eqn:init_sig}--\ref{eqn:init_omg} below and surrounding
discussion). The wobbling generates spurious time variability that increases with increasing $q$; therefore we drop the indirect potential in all \texttt{ZEUS90} simulations with $q\geq 1\times 10^{-3}$
(apparently
\citealt{lubow99} also dropped the indirect potential in their
simulations with \texttt{ZEUS}; see also \citealt{Zhu11} who found 
that their planet-disk simulations were not sensitive to the indirect term).
Running in double precision on 128 cores in
parallel, \texttt{ZEUS90} takes 26 seconds to run per planetary orbit
for $(q,\alpha,h/r) = (10^{-3},10^{-2},0.05)$.

\subsection{Numerical Setup} \label{sec:setup}
For our parameter study we vary:
\begin{itemize}
\item $q$ from $10^{-4}$ to $10^{-2}$,
\item $\alpha$ from $10^{-3}$ to $10^{-1}$, and
\item $h/r$ from $0.04$ to $0.1$.
\end{itemize}

Other properties of our simulations are as follows.

\subsubsection{Initial and boundary conditions} \label{sec:boundary}
\begin{figure}[]
\includegraphics[width=0.99\columnwidth]{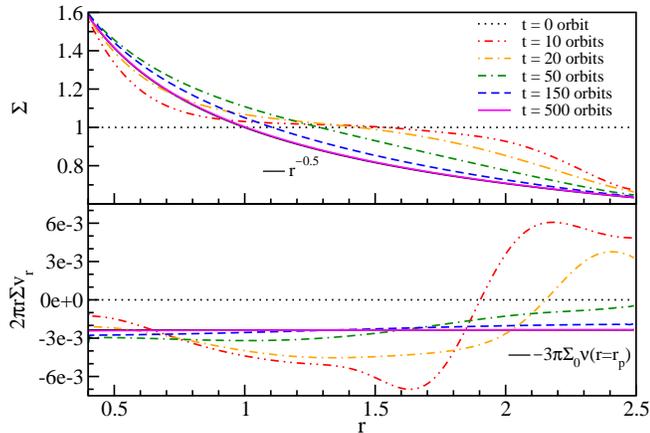}
\caption{Viscous relaxation to steady-state accretion in a
  disk with $(q,\alpha,h/r) = (0, 0.1, 0.05)$.  At $t=0$, we set
  $\Sigma= 1$ and $v_r = 0$ except at the boundaries, where conditions
  are given by equations (\ref{eqn:init_sig})--(\ref{eqn:init_omg}).
  Black solid lines denote the steady-state density profile and
  accretion rate to which \texttt{PEnGUIn} correctly relaxes over a
  viscous timescale.}
\label{fig:relax}
\end{figure}

Our simulation domain spans 0 to $2\pi$ in azimuth, and from $r_{\rm
  in} = 0.4$ to $r_{\rm out} = 2.5$ in radius (in units where the
planet-star separation $r_{\rm p} = 1$).  Initial conditions
correspond to a steady-state accretion disk having constant $\alpha$,
constant $h/r$, and a rotation curve modified by the radial
pressure gradient:
\begin{align}
\label{eqn:init_sig}
\Sigma &= \Sigma_0 \, (r/r_{\rm p})^{-1/2} \,,\\
\label{eqn:init_vr}
v_{r} &= -\frac{3}{2}\, \alpha\, \left(\frac{h}{r}\right)^2\sqrt{\frac{G(M_\ast+M_{\rm p})}{r}}\,,\\
\label{eqn:init_omg}
\Omega &= \sqrt{1-\frac{1}{2}\left(\frac{h}{r}\right)^2}\sqrt{\frac{G(M_\ast+M_{\rm p})}{r^3}}\,,
\end{align}
with $\Sigma_0 = 1$ (we could have chosen any value for $\Sigma_0$ because we do not calculate the gravity of the disk---the disk exerts no gravitational backreaction on the planet nor does it self-gravitate).
At both inner and outer radial boundaries, we fix
$\Sigma$, $v_r$ and $\Omega$ to their values determined by the above
equations. These fixed boundary conditions ensure a steady inflow of
mass across the simulation domain --- as is appropriate for real accretion
disks.\footnote{By contrast, many
  popular codes for planet-disk simulations (e.g., \texttt{FARGO}) default to a
  zero-inflow solution; for a compilation of codes from the community,
  see, e.g., \citet{devalborro06}.}
Figure \ref{fig:relax} illustrates how our boundary conditions
enable a planet-less disk to relax over a viscous diffusion timescale to the
equilibrium profile described by equations
(\ref{eqn:init_sig})--(\ref{eqn:init_omg}).

Ideally the radial boundaries should be placed far enough away that
waves launched from the planet damp before they reach the edges of the
domain.
\citet{GR2001} calculated that nonlinear steepening of waves causes
them to dissipate over lengthscales of $\sim$$3h$ 
(for $q=10^{-3}$ and $h/r=0.05$; the damping length scales as $q^{-0.4}$ and $(h/r)^{2.2}$). 
Our outer radial boundary of $r_{\rm out}=2.5$ ($\sim$15--40$h$ away
from the planet, depending on our choice for $h/r$) is distant enough
that outward-propagating waves largely dissipate within the domain.
For our inner radial boundary of $r_{\rm in}=0.4$ ($\sim$6--15$h$ away
from the planet), the situation is more marginal; depending on the
simulation, waves are still present at our disk inner edge. However,
the main focus of our paper is the surface density deep within the
gap, and we have verified that this quantity changes by no more than
$\sim$10\% as we shrink $r_{\rm in}$ from 0.4 to 0.2. Thus we opt for
the larger boundary to keep code timesteps longer. For simplicity we
eschew wave-killing zones (cf.~\citealt{devalborro06}).  In practice,
the Godunov-type scheme used by \texttt{PEnGUIn} is effective at absorbing
waves at fixed boundaries, even more so than using wave-killing zones
(Zhaohuan Zhu 2013, personal communication).

To avoid strong shocks at the beginning of the simulation, the planet
mass is ramped from zero to its assigned value over an initial
``warm-up'' phase. In \texttt{PEnGUIn}, $M_{\rm p}$ increases according to
${M_{\rm p}(t)}/{M_\ast} = q \sin^2 [(\Omega_{\rm p}t /20) (10^{-3}/q) ]$.
For $q=10^{-3}$, this takes $5$ orbits.  In \texttt{ZEUS90}, the planet
mass grows linearly from zero to its desired value in 1 orbit. 
Both warm-up schemes proved stable.

\subsubsection{Grid resolution}\label{sec:resolution}

\begin{figure}[]
\includegraphics[width=0.99\columnwidth]{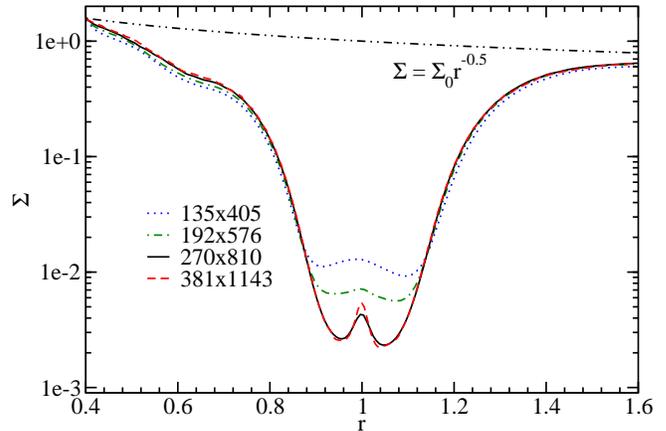}
\caption{Convergence of gap profile with grid resolution for $(q,\alpha,h/r) =
  (10^{-3}, 10^{-3}, 0.05)$ using \texttt{PEnGUIn}. 
The dot-dot-dashed curve
  represents the initial density profile, equal to the density profile
in the absence of the planet (equation \ref{eqn:init_sig}). The surface
  density $\Sigma$ plotted here is azimuthally averaged. For \texttt{PEnGUIn} science 
  runs, we adopt $270 \,(r)\times 810 \,(\phi)$ for $h/r=0.05$, and adjust the cell
  size to scale with $h/r$ (see \S\ref{sec:resolution}).}
\label{fig:res}
\end{figure}

Our grid spacings are logarithmic in radius and uniform in azimuth.
For $h/r = 0.05$, the resolution is $270 \,(r) \times 810 \,(\phi)$ for
\texttt{PEnGUIn} and $256\times 864$ for \texttt{ZEUS90} 
(the latter choice yields square grid cells). 
Figure \ref{fig:res} attests that
gap surface densities have largely converged at our standard
resolution. We scale our grid cell size with $h/r$, i.e., with
sound speed $c_{\rm s}$, so that
sound waves of a given frequency are equally well resolved
between simulations. Cold disks with small $h/r$ are especially
costly, which is why we do not vary $h/r$ below 0.04.
Code timesteps scale with grid cell sizes according
to the Courant-Friedrichs-Lewy condition, with the Courant
number chosen to be 0.5 for \texttt{PEnGUIn}
and 0.4 for \texttt{ZEUS90}.

\subsubsection{Smoothing length $r_s$} \label{sec:smooth}
For both \texttt{PEnGUIn} and \texttt{ZEUS90}, the planetary
potential's softening length is fixed at $r_s = 0.028 r_{\rm p}$ or
about 4 local grid cell lengths.  Equivalently, $r_s = 0.56 h$ for
$h/r=0.05$, and $r_s = 0.25$ Hill radii $R_{\rm H}$ for $q = 10^{-3}$.
Any choice for $r_s \sim h$ or $r_s \sim R_{\rm H}$ seems reasonable
insofar as our 2D treatment of the gas dynamics must break down at
distances from the planet less than the vertical thickness of the
disk, and because the planet's mass may, in reality, be distributed
over a distended envelope or circumplanetary disk.  Tests
with \texttt{PEnGUIn} at $(q,\alpha,h/r) = (10^{-3}, 0.1, 0.05)$
revealed that $r_s \lesssim 0.4h$ caused the surface density to
converge substantially more slowly with time. Specifically, for the
aforementioned parameters and $r_s$
too small, the gap deepened rapidly, overshot its equilibrium value,
and took thousands of orbits to approach a
steady state. By contrast, for $r_s = 0.56 h$, the surface density
equilibrated in a mere $\sim$30 orbits at our standard resolution,
with higher grid resolutions yielding similar results.

According to \citet{Muller12}, our choice of smoothing length 
yields a 2D gravitational force that matches the vertically averaged, 3D force
to within 10\% at a distance $\gtrsim2 h$ away from the planet.

\subsubsection{$\Sigma_{\rm gap}$ and convergence with time}\label{sec:metric}
\begin{figure*}[]
\includegraphics[width=1.99\columnwidth]{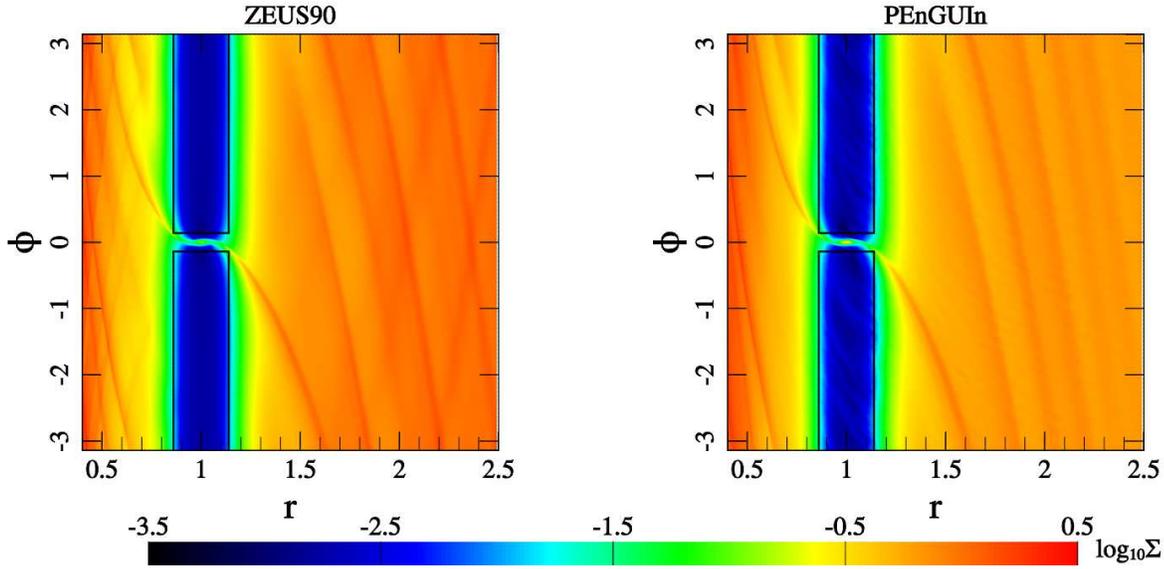}
\caption{Snapshots of simulations with $(q,\alpha,h/r) = (10^{-3},
  10^{-3}, 0.05)$. \texttt{PEnGUIn}'s snapshot is taken at $t=2\times
  10^4 P_{\rm p}$ while \texttt{ZEUS90}'s is taken at $t=1\times 10^4
  P_{\rm p}$. Overall the two codes agree well on the shape and depth
  of the gap. \texttt{ZEUS90} has more trouble converging to the
  desired outer boundary condition; $\Sigma$ at $r = 2.5$ deviates
  from that imposed by equation (\ref{eqn:init_sig}) by up to
  $\sim$50\%. 
Note that \texttt{PEnGUIn} does not have the problem in the
    outer disk that \texttt{ZEUS90} does, and moreover succeeds in
    resolving fine streamers (``filaments'') within the gap.  The black rectangles
  indicate the area over which $\Sigma_{\rm gap}$ is averaged.}
\label{fig:snapshots}
\end{figure*}

\begin{figure*}[]
\includegraphics[width=1.99\columnwidth]{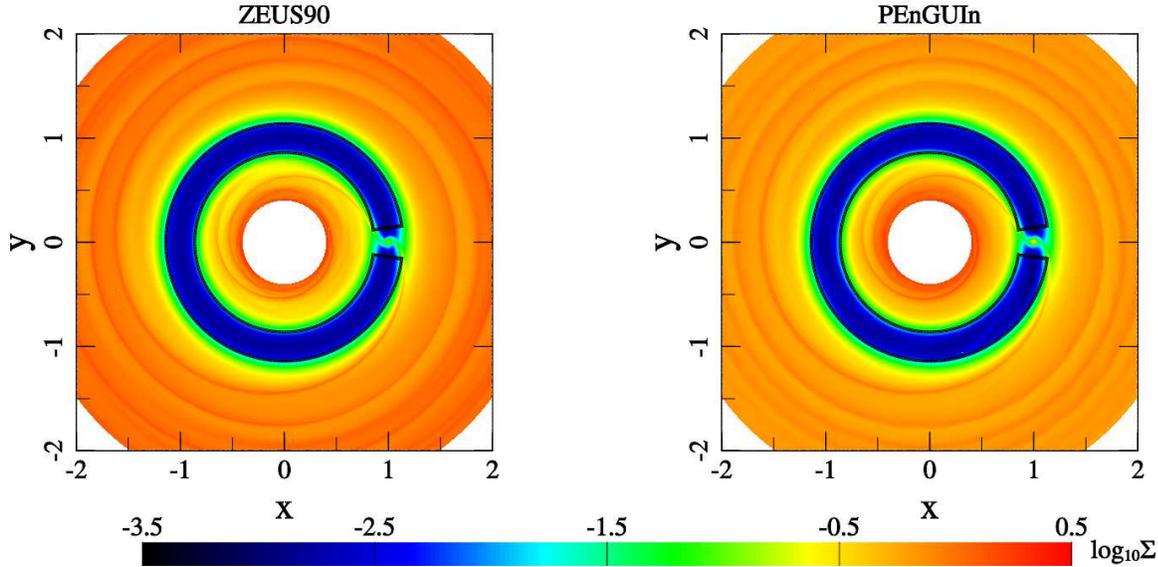}
\caption{Cartesian version of Figure \ref{fig:snapshots}.}
\label{fig:snapshots_xy}
\end{figure*}

Our metric for gap depth is the space- and time-averaged surface
density $\Sigma_{\rm gap}$ in the planet's co-rotation region,
normalized to $\Sigma_0 = 1$ (the surface density at $r = r_{\rm p} = 1$ in the absence of the
planet).  As judged from snapshots like those shown in Figures
\ref{fig:snapshots} and \ref{fig:snapshots_xy},
the annulus spanning $r= r_{\rm p} -\Delta$ to
$r_{\rm p} + \Delta$ with $\Delta \equiv 2 \max (R_{\rm H}, h)$,
excised from $\phi = \phi_{\rm p} -\Delta/r_{\rm p}$ to $\phi_{\rm p}
+ \Delta/r_{\rm p}$, is visibly depleted and reasonably uniform. For
most simulations, this is the area over which we average $\Sigma$ to
calculate $\Sigma_{\rm gap}$.

In a few cases the outer edge of the gap is visibly eccentric (see
\S\ref{sec:high_q} for more discussion), and the circular annulus we
have defined above becomes contaminated with non-gap material and is
no longer suitable for measuring $\Sigma_{\rm gap}$.  In these cases,
we keep the circular inner gap edge and the azimuthal excision as
defined above, but approximate the outer gap edge with an ellipse
having semimajor axis $r_{\rm p} + \Delta$, and an eccentricity and
apsidal orientation estimated by eye from snapshots (for a sampling,
jump to Figures \ref{fig:eccentric} and \ref{fig:eccentric_xy}).

\begin{figure}[]
\includegraphics[width=0.99\columnwidth]{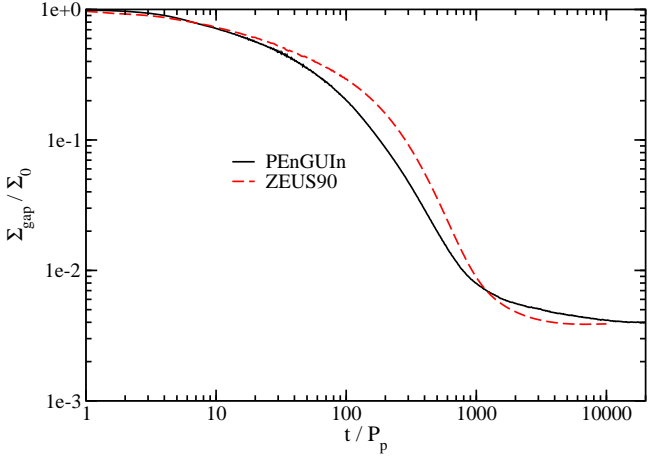}
\caption{Convergence of $\Sigma_{\rm gap}$ with time for
  $(q,\alpha,h/r) = (10^{-3}, 10^{-3}, 0.05)$. For these parameters,
  the viscous timescale is formally $r_{\rm p}^2/\nu \sim 6 \times
  10^4$ planetary orbits.}
\label{fig:time}
\end{figure}

Each simulation runs until $\Sigma_{\rm gap}$ appears to have
converged in time; see Figure \ref{fig:time} for an example.  The time
required to reach convergence scales approximately as the viscous
timescale $r_{\rm p}^{2}/\nu$, shortening somewhat with larger $q$.
Each value of $\Sigma_{\rm gap}$ that we report in Table
\ref{tab:tab1} is averaged over a time interval that starts at time
$t_{\rm conv}$ near the end of the simulation, and that lasts for
duration $\Delta t$. For many models, there is actually no need to
time-average because the time variability in $\Sigma_{\rm gap}$ is
fractionally small (less than a few percent).  However, some models
exhibit greater variability, particularly for $q$ approaching
$10^{-2}$ or $h/r = 0.1$.  The fluctuations appear periodic, with
periods ranging from 0.5--1 $P_{\rm p}$ and amplitudes up to order
unity.  For these more strongly time-variable cases, we also record in
Table \ref{tab:tab1} the maximum and minimum values of $\Sigma_{\rm
  gap}$ that occur during the averaging interval.

\begin{deluxetable*}{cccrrccccccccc}
\tabletypesize{\footnotesize}
\tablecolumns{14}
\tablewidth{0pc}
\tablecaption{Simulated Gap Depths}
\setlength{\tabcolsep}{0.07in}
\tablehead{
\colhead{} & \colhead{} & \colhead{} &
\colhead{\texttt{PEnGUIn}}  & \colhead{\texttt{ZEUS90}} & \colhead{} &
\multicolumn{2}{c}{\texttt{PEnGUIn}} & \colhead{} &
\multicolumn{2}{c}{\texttt{ZEUS90}} &
\colhead{} & \colhead{\texttt{PEnGUIn}} & \colhead{\texttt{ZEUS90}} \\
\cline{4-5} \cline{7-8} \cline{10-11} \cline{13-14}\\
\colhead {$q$} & \colhead {$\alpha$} & \colhead {$h/r$} &
\multicolumn{2}{c}{ $\Sigma_{\rm gap}\tablenotemark{a}~(\Sigma_0)$}     &  \colhead{} &
\colhead{ $t_{\rm conv}\tablenotemark{b}~(P_{\rm p})$ } & 
\colhead{ $\Delta t\tablenotemark{b}~(P_{\rm p})$ } &  \colhead{} &
\colhead{ $t_{\rm conv}~(P_{\rm p})$ } &
\colhead{ $\Delta t~(P_{\rm p})$ } &
			\colhead{} & \multicolumn{2}{c}{Comments}
}
\startdata
$1\times 10^{-4}$ & $10^{-3}$ & 0.05 & $4.6\times 10^{-1}$ & $4.9\times 10^{-1}$ &
&  20000 & 10 &
&  10000 & 10 &
&        & d \\
$2\times 10^{-4}$ & $10^{-3}$ & 0.05 & $1.9\times 10^{-1}$ & $2.0\times 10^{-1}$ &
& 20000 & 10 &
& 10000 & 10 &
&       & d \\
$5\times 10^{-4}$ & $10^{-3}$ & 0.05 & $3.0\times 10^{-2}$ & $3.1\times 10^{-2}$ &
& 20000 & 10 &
& 10000 & 10 &
&       & d \\
$1\times 10^{-3}$ & $10^{-3}$ & 0.05 & $4.0\times 10^{-3}$ & $3.9\times 10^{-3}$ &
& 20000 & 10 &
& 10000 & 10 & 
&       &  \\
$2\times 10^{-3}$ & $10^{-3}$ & 0.05 & $8.5^{+0.4}_{-0.3}\times 10^{-4}$ & $1.2^{+0.1}_{-0.2}\times 10^{-3}$ &
&  6000 & 10 &
&  6000 & 10 &
&       & e  \\
$5\times 10^{-3}$ & $10^{-3}$ & 0.05 & $2.1^{+0.2}_{-0.2}\times 10^{-4}$ & 2.6$^{+0.5}_{-0.8}\times 10^{-4}$ &
&  6000 & 10 &
&  6000 & 10 &
&  f    & e  \\
$1\times 10^{-2}$ & $10^{-3}$ & 0.05 & $1.7^{+0.4}_{-0.5}\times 10^{-4}$ & $1.4^{+0.7}_{-0.5}\times 10^{-4}$ &
&  6000 & 10 &
&  6000 & 10 &
&  f    &   \\
\tableline
$1\times 10^{-3}$ & $10^{-2}$ & 0.05 & $1.4\times 10^{-1}$ & $1.5\times 10^{-1}$ &
&  2000 & 10 &
&  2000 & 10 &
&       &  \\
$2\times 10^{-3}$ & $10^{-2}$ & 0.05 & $2.7\times 10^{-2}$ & $2.8\times 10^{-2}$ &
&  2000 & 10 &
&  2000 & 10  &
&       &  \\
$5\times 10^{-3}$ & $10^{-2}$ & 0.05 & $5.5^{+0.9}_{-1.0}\times 10^{-3}$ & $4.6^{+2.6}_{-1.8}\times 10^{-3}$ &
&  2000 & 10 &
&  2000 & 10 &
&       &  \\
$1\times 10^{-2}$ & $10^{-2}$ & 0.05 & $2.0^{+0.6}_{-0.8}\times 10^{-3}$ & $2.7^{+0.6}_{-0.4}\times 10^{-3}$ &
&  2000 & 10 &
&  2000 & 10 &
&  e    &   \\
\tableline
$1\times 10^{-3}$ & $10^{-1}$ & 0.05 & $6.1\times 10^{-1}$ & $7.2\times 10^{-1}$ &
&   300 & 10 &
&   300 & 10 &
&       &   \\
$2\times 10^{-3}$ & $10^{-1}$ & 0.05 & $3.5\times 10^{-1}$ & $4.9\times 10^{-1}$ &
&   300 & 10 &
&   300 & 10 &
&       &   \\
$5\times 10^{-3}$ & $10^{-1}$ & 0.05 & $9.8\times 10^{-2}$ & $1.7\times 10^{-1}$ &
&   300 & 10 &
&   300 & 10 &
&       &  \\
$1\times 10^{-2}$ & $10^{-1}$ & 0.05 & $3.8\times 10^{-2}$ & $4.9^{+0.2}_{-0.2}\times 10^{-2}$ &
&   300 & 10 &
&   300 & 10 &
&       &    \\
\tableline
$1\times 10^{-3}$ & $10^{-3}$ & 0.1  & $2.2\times 10^{-1}$ & $2.4^{+0.1}_{-0.1}\times 10^{-1}$ &
&  7000 & 10 &
&  7000 & 10 &
&       &   \\
$5\times 10^{-3}$ & $10^{-3}$ & 0.1  & $1.5^{+0.5}_{-0.1}\times 10^{-2}$ & $1.6^{+0.1}_{-0.1}\times 10^{-2}$ &
&  6000 & 1000 &
&  7000 & 10   &
&  c    &  \\
$1\times 10^{-2}$ & $10^{-3}$ & 0.1  & $8.4^{+0.7}_{-0.6}\times 10^{-3}$ & $1.0^{+0.1}_{-0.1}\times 10^{-2}$ &
&  6900 & 100 &
&  7000 & 10  &
&  c    &  \\
\tableline
$5\times 10^{-4}$ & $10^{-3}$ & 0.04 & $6.0\times 10^{-3}$ & $6.5\times 10^{-3}$ &

& 17000 & 10 &
& 10000 & 10 &
&       & d \\
$2\times 10^{-3}$ & $10^{-3}$ & 0.04 & $2.7^{+0.2}_{-0.2}\times 10^{-4}$ & $2.9^{+0.6}_{-0.6}\times 10^{-4}$ &
& 10000 & 10 &
& 5000  & 10 &
&       & e \\
\tableline
$2\times 10^{-3}$ & $10^{-2}$ & 0.04 & $7.9^{+1.1}_{-0.9}\times 10^{-3}$ & $7.0^{+0.4}_{-0.5}\times 10^{-3}$ &
& 4000  & 10 &
& 4000  & 10 &
&       &  \\
\enddata \label{tab:tab1}

\tablenotetext{a}{Averaged over time and over a partial annulus
  centered on the planet, as defined in \S\ref{sec:metric} and
  delineated in Figures \ref{fig:snapshots} and \ref{fig:snapshots_xy}.
  For visibly eccentric
  outer disks (see Comments column), the outer edge of the measurement
  annulus is made eccentric to conform to the gap shape (e.g., Figures
  \ref{fig:eccentric} and \ref{fig:eccentric_xy}). Maximum and
  minimum values of $\Sigma_{\rm gap}$ are given for runs for which
  these values deviate from the time-averaged value by more than a
  percent. All surface densities are in units where $\Sigma(r=1)=1$ in
  a steadily accreting, planet-less disk.}

\tablenotetext{b}{The time $t_{\rm conv}$ is taken near the end of a
  simulation, when $\Sigma_{\rm gap}$ appears to have nearly converged
  to its steady-state value (see Figure \ref{fig:time}). Formally
  $t_{\rm conv}$ marks the beginning of the time interval, of duration
  $\Delta t$, over which $\Sigma_{\rm gap}$ is averaged. All times
  listed are in units of the planetary orbital period, $P_{\rm p}$. }

\tablenotetext{c}{Highly unsteady outer gap edge (e.g., Figure
  \ref{fig:unsteady}) and long-term time variability. A longer $\Delta
  t$ is chosen to capture the variability.}

\tablenotetext{d}{Indirect potential included (\S\ref{sec:zeus}).}

\tablenotetext{e}{Eccentric outer disk (e.g., Figures
  \ref{fig:eccentric} and \ref{fig:eccentric_xy}). Measured
  eccentricities are $\sim$0.10--0.15 and apsidal precession periods
  are $\sim$300--$600 P_p$.}

\tablenotetext{f}{An eccentric outer disk is observed at $t\sim 1000
  P_{\rm p}$, but the eccentricity damps away by $t_{\rm conv}$. The damping
  is probably artificial (\S\ref{sec:high_q}).}

\end{deluxetable*}

\section{RESULTS \label{sec:result}}
In \S\ref{sec:empirical}, we obtain empirical scalings for gap depths
at $10^{-4} \leq q \leq 5 \times 10^{-3}$; in \S\ref{sec:high_q}
we repeat for $5 \times 10^{-3} \leq q \leq 10^{-2}$
and discuss qualitatively the new dynamical phenomena
that appear at these highest companion masses;
and in \S\ref{sec:compare} we highlight some of the differences
between \texttt{PEnGUIn} and \texttt{ZEUS90}.

\subsection{Gap Depth Scalings for $10^{-4} \leq q \leq 5 \times 10^{-3}$}
\label{sec:empirical}

\begin{figure}[]
\includegraphics[width=0.99\columnwidth]{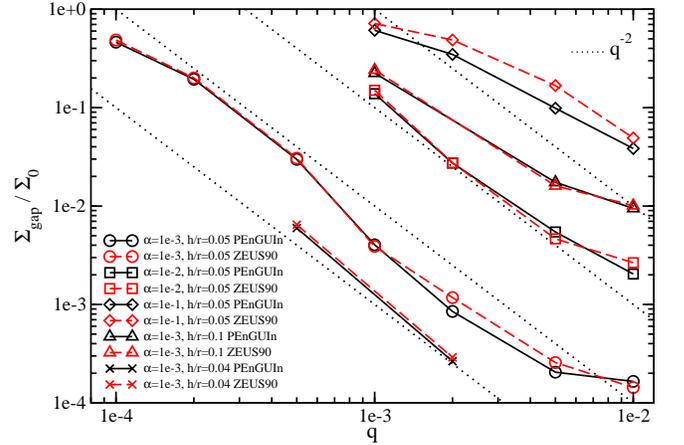}
\caption{$\Sigma_{\rm gap}$ vs.~$q$. Black dotted lines indicate constant power-law slopes of $-2$, and are shown for reference only. The power-law slopes approximately equal $-2$ for $q < 5 \times 10^{-3}$, and flatten to $-1$ for higher $q$. 
For formal power-law fits, see the main text.
}
\label{fig:q_plot}
\end{figure}

\begin{figure}[]
\includegraphics[width=0.99\columnwidth]{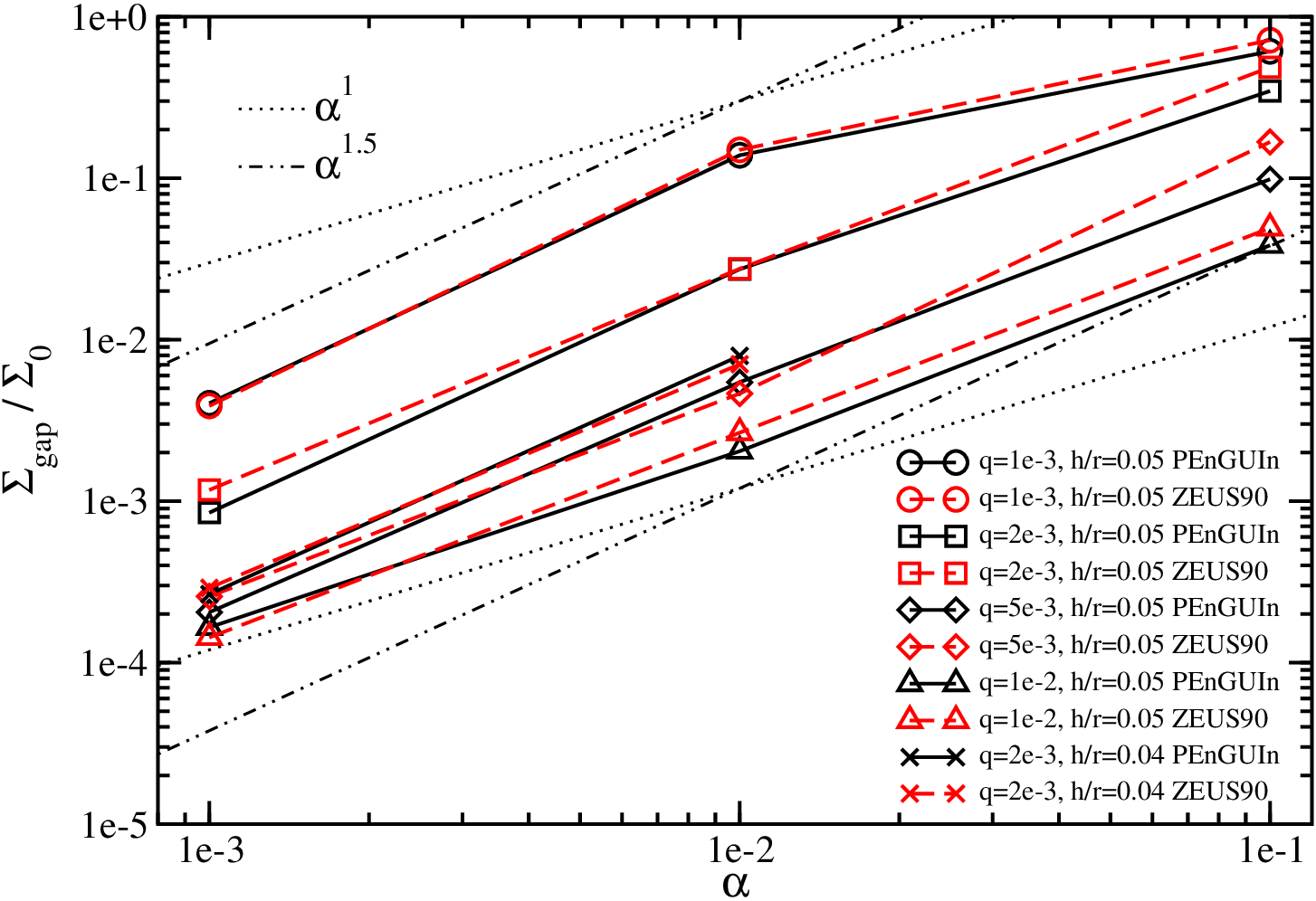}
\caption{$\Sigma_{\rm gap}$ vs.~$\alpha$. Dotted and dot-dot-dashed
  lines indicate power-law slopes of 1 and 1.5, bracketing the range
  exhibited by the data. For formal power-law fits, see main text.  }
\label{fig:a_plot}
\end{figure}

\begin{figure}[]
\includegraphics[width=0.99\columnwidth]{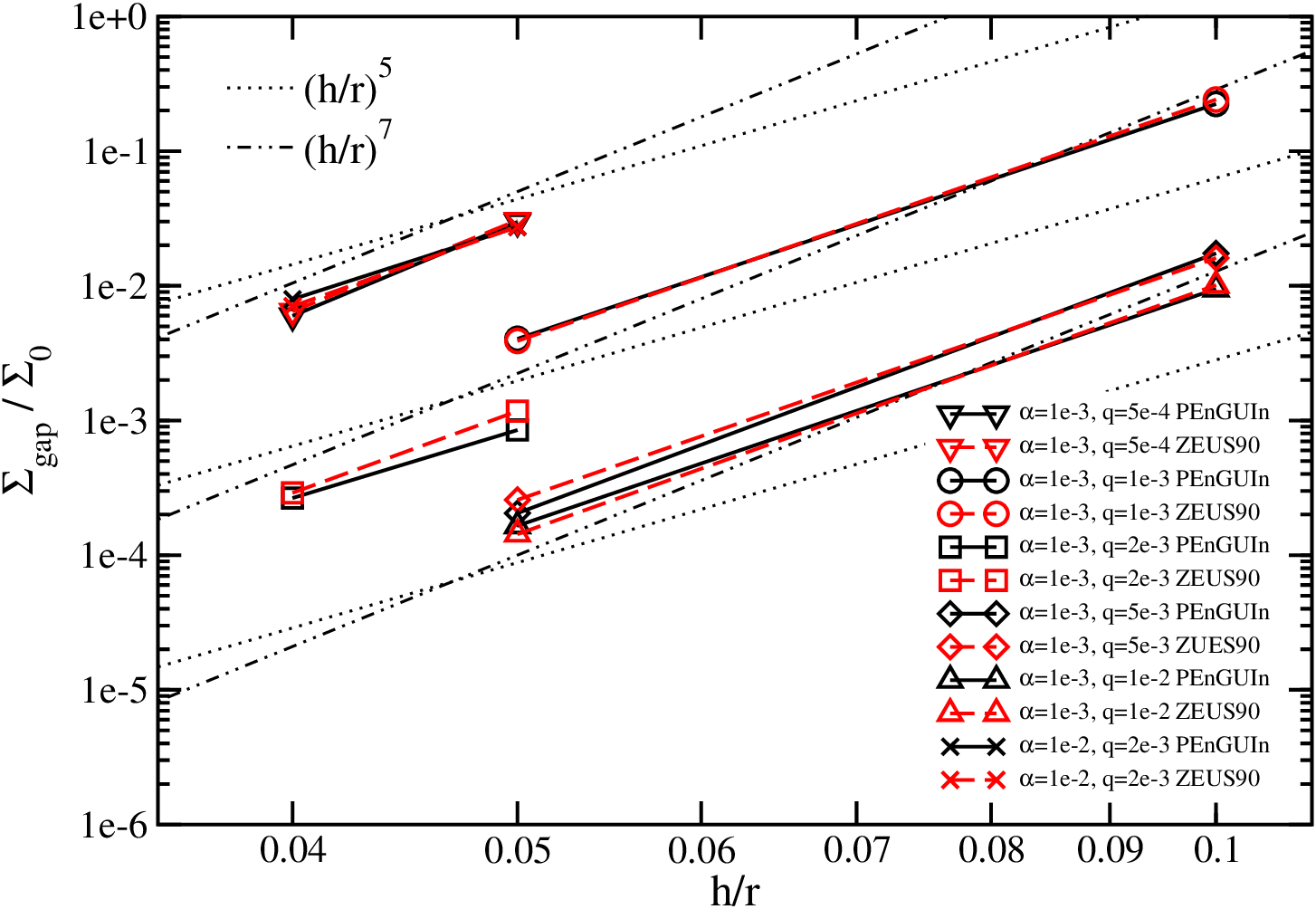}
\caption{$\Sigma_{\rm gap}$ vs.~$h/r$. Dotted and dot-dot-dashed lines
  indicate power-law slopes of 5 and 7, bracketing the range exhibited
  by the data. For formal power-law fits, see main text.  }
\label{fig:h_plot}
\end{figure}

Gap depths $\Sigma_{\rm gap}(\Sigma_0)$ are recorded in
Table~\ref{tab:tab1} and plotted against each of the parameters $q$,
$\alpha$, and $h/r$ in Figures \ref{fig:q_plot}, \ref{fig:a_plot}, and
\ref{fig:h_plot}, respectively.  Overall the agreement between the two
codes, which utilize completely different algorithms, is remarkably good.

Figure \ref{fig:q_plot} attests that
for $q \lesssim 5 \times 10^{-3}$, gap depths scale roughly as
$q^{-2}$ --- as our analytic scaling (\ref{eqn:scaling}) predicts.
For $q \gtrsim 5 \times 10^{-3}$, the curves flatten somewhat (more on
the behavior at large $q$ in \S\ref{sec:high_q}).  In Figure
\ref{fig:a_plot}, $\Sigma_{\rm gap}$ appears to scale with $\alpha$ to
a power between 1 and 1.5.  For comparison, our analytic scaling
(\ref{eqn:scaling}) predicts $\Sigma_{\rm gap} \propto \alpha^1$.  The
empirical dependence on $h/r$ is similarly steeper than the analytic
dependence: equation (\ref{eqn:scaling}) predicts that $\Sigma_{\rm
  gap} \propto (h/r)^5$ whereas Figure \ref{fig:h_plot} shows that the
power-law indices vary between 5 and 7. 

We obtain a ``best-fit'' power-law relation by minimizing the function
$y = \sum [\ln D q^A \alpha^B (h/r)^C - \ln \Sigma_{\rm gap}]^2/N$
over the parameters $(A,B,C,D)$. The sum is performed over $N$
data points, excluding the discrepant runs at $q=0.01$ and runs for
which $\Sigma_{\rm gap}/\Sigma_0 > 0.2$ (i.e., runs for which gaps
hardly open).  With these exclusions, there are $N=13$ data points
from \texttt{PEnGUIn}, best fitted by
\begin{align}
\nonumber
10^{-4} \leq \, q \leq& \, 5 \times 10^{-3} \,: \\
\label{eqn:bestfit_p}
\Sigma_{\rm gap}/\Sigma_0 =& \, 0.14 \left( \frac{q}{10^{-3}} \right)^{-2.16} \left( \frac{\alpha}{10^{-2}} \right)^{1.41} \left( \frac{h/r}{0.05} \right)^{6.61} \,.
\end{align}
The $N=13$ points from \texttt{ZEUS90} are best described by a very similar formula:
\begin{align}
\nonumber
10^{-4} \leq \, q \leq& \, 5 \times 10^{-3} \,: \\
\label{eqn:bestfit_z}
\Sigma_{\rm gap}/\Sigma_0 =& \, 0.15 \left( \frac{q}{10^{-3}} \right)^{-2.12} \left( \frac{\alpha}{10^{-2}} \right)^{1.42} \left( \frac{h/r}{0.05} \right)^{6.45} \,.
\end{align}
Both of these relations fit their respective data to typically better
than 20\%; the largest deviation in the \texttt{PEnGUIn} fit is 40\%,
corresponding to $(q,\alpha,h/r) = (2\times 10^{-3}, 10^{-3}, 0.05)$, and for
\texttt{ZEUS90} the largest deviation is 50\%, corresponding to
$(q,\alpha,h/r) = (10^{-3}, 10^{-3}, 0.05)$.

Because our two codes agree so well, and because the fits are good, we
are confident the deviations between our empirical scaling (say
equation \ref{eqn:bestfit_p} from \texttt{PEnGUIn}) and our analytic
scaling (\ref{eqn:scaling}) are real and reflect physical
  effects not captured by our analytic scaling.  And because our
analytic scaling (\ref{eqn:scaling}) matches exactly the scalings
found numerically by \citet{duffell13} at low $q \lesssim 10^{-4}$ ---
whereas our empirical relation (\ref{eqn:bestfit_p}) applies to high
$q \gtrsim 10^{-4}$ --- these physical effects manifest for giant
(Jupiter-like) planets, not lower-mass (Neptune-like) planets.\footnote{Another difference between our simulations and theirs is
  that we mandate a steady $\dot{M} \neq 0$ across our entire domain,
  whereas they adopt (as appears customary for work in this field)
  wave-killing zones that effectively result in nearly zero-inflow boundary
  conditions. We have verified, however, that this difference does not
  matter for $\Sigma_{\rm gap}$; we implemented zero-inflow boundaries
  in a few runs with \texttt{PEnGUIn} and found results for
  $\Sigma_{\rm gap}$ that matched those with our standard accreting
  boundaries to better than 1\%.}
We have not elucidated what this physics is,
although there might be some clues from the gap behavior
at the very highest values of $q$ we tested, as discussed in
\S\ref{sec:high_q}.

At the same time, we emphasize that the deviations between
(\ref{eqn:scaling}) and (\ref{eqn:bestfit_p}), though (probably) real,
are not large. If we insist on fitting the \texttt{PEnGUIn} data using
(\ref{eqn:scaling}) --- i.e., if we fix $(A, B, C) = (-2, 1, 5)$ and
allow only the coefficient $D$ to float --- then the data deviate from
(\ref{eqn:scaling}) by typically a factor of 2, and at most a factor
of 3. Thus the physical effects not captured by our analytic scaling, whatever they are, do not lead to
order-of-magnitude changes in gap depth, at least over the range of
parameters tested.


\subsection{Behavior of Gaps at High $q \gtrsim 5\times 10^{-3}$}\label{sec:high_q}

The dependence of $\Sigma_{\rm gap}$ on $q$ flattens at the highest
values of $q$ considered (Figure \ref{fig:q_plot}).
Fitting the $N=8$ points from \texttt{PEnGUIn} for which
$q \geq 5 \times 10^{-3}$ and $\Sigma_{\rm gap}/\Sigma_0 < 0.2$
yields:
\begin{align}
\nonumber
10^{-2} \geq \, q \geq& \, 5 \times 10^{-3} \,: \\
\label{eqn:bestfit_highq_p}
\Sigma_{\rm gap}/\Sigma_0 =& \, 4.7 \times 10^{-3} \left( \frac{q}{5 \times 10^{-3}} \right)^{-1.00} \left( \frac{\alpha}{10^{-2}} \right)^{1.26} \left( \frac{h/r}{0.05} \right)^{6.12} \,.
\end{align}
Similarly for the $N=8$ points from \texttt{ZEUS90} we obtain:
\begin{align} 
\nonumber
10^{-2} \geq \, q \geq& \, 5 \times 10^{-3} \,: \\
\label{eqn:bestfit_highq_z}
\Sigma_{\rm gap}/\Sigma_0 =& \, 5.6 \times 10^{-3} \left( \frac{q}{5 \times 10^{-3}} \right)^{-1.02} \left( \frac{\alpha}{10^{-2}} \right)^{1.34} \left( \frac{h/r}{0.05} \right)^{6.12} \,.
\end{align}
Although at first glance one might attribute the flattening of the
trend of $\Sigma_{\rm gap}$ with $q$ to the onset of strong shocks,
we do not believe this is the
correct interpretation.  In the strong-shock regime, where
disturbances excited by the planet are non-linear at launch, 
the torque exerted on the disk by the planet
 scales as $q^1 (h/r)^0$ (e.g.,
\citealt{HQ11}, their section 2.3).\footnote{This scaling can
be seen by replacing $h$ with $R_{\rm H}$ in equation
(\ref{eqn:lindblad}); the torque cut-off distance generally equals
$\max (R_{\rm H}, h)$, which in the strong-shock limit equals $R_{\rm
  H}$.} Then the same arguments in \S\ref{sec:scaling} yield
$\Sigma_{\rm gap} \propto q^{-1} (h/r)^2$. This analytic relation
reproduces the scaling index for $q$ given by our empirical relations (\ref{eqn:bestfit_highq_p}) and
(\ref{eqn:bestfit_highq_z}), but fails to reproduce
the empirical scaling index for $h/r$.  Furthermore, the
flattening begins at an apparently ``universal'' $q$-value of 
$\sim$$5 \times 10^{-3}$ that is independent of $h/r$, whereas in the
strong-shock interpretation, the critical $q$-value should scale as
$(h/r)^3$ (i.e., the expected critical $q$ is given by the so-called
thermal mass).

\begin{figure*}[]
\includegraphics[width=1.99\columnwidth]{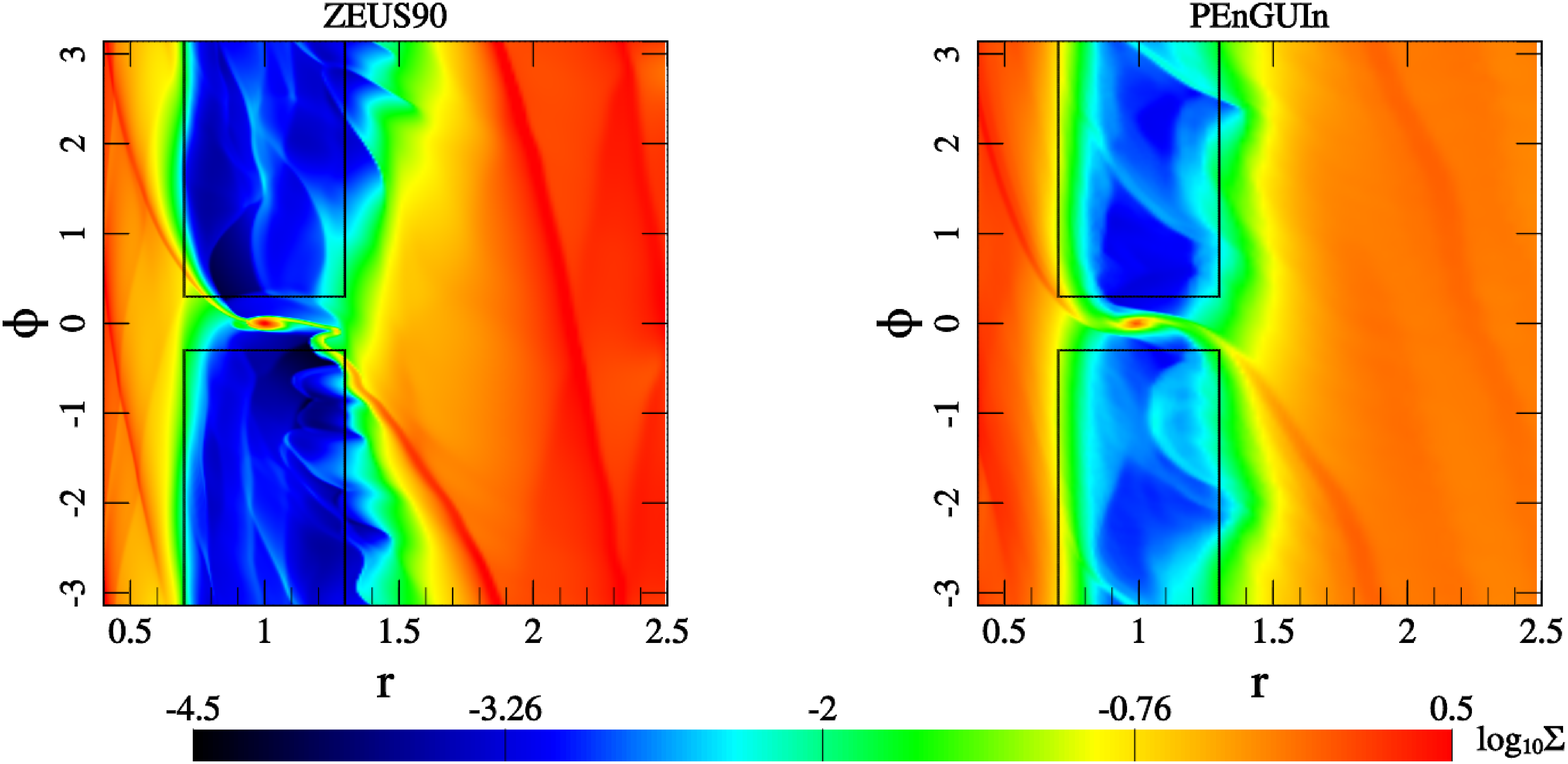}
\caption{Two different examples at high $q$ of unsteady gap edges
and streamers filling gaps. The \texttt{ZEUS90} snapshot
is for $(q,\alpha,h/r) = (0.01,0.01,0.05)$ and the
\texttt{PEnGUIn} snapshot is for $(q,\alpha,h/r) = (0.01, 0.001, 0.1)$.
}
\label{fig:unsteady}
\end{figure*}

We do not have an explanation for the flatter slope of $-1$ at high
$q$. We speculate that it might be caused by the most massive companions
causing material at the gap edge to ``leak'' into the gap. The most
massive planets disturb the gap edge so strongly that local
instabilities send streamers of gas into the gap. These streamers,
which \citet{devalborro06} called ``filaments'', are prominent in the
high-$q$ snapshots in Figure \ref{fig:unsteady} (and can be seen even
at $q=10^{-3}$ in the \texttt{PEnGUIn} snapshot in Figure
\ref{fig:snapshots}). 
The filaments appear to originate from unsteady
structures along gap edges; similar structures were seen by, e.g., Kley
\& Dirksen (\citeyear{KD2006}, their figures 1 and 7).

\begin{figure*}[]
\includegraphics[width=1.99\columnwidth]{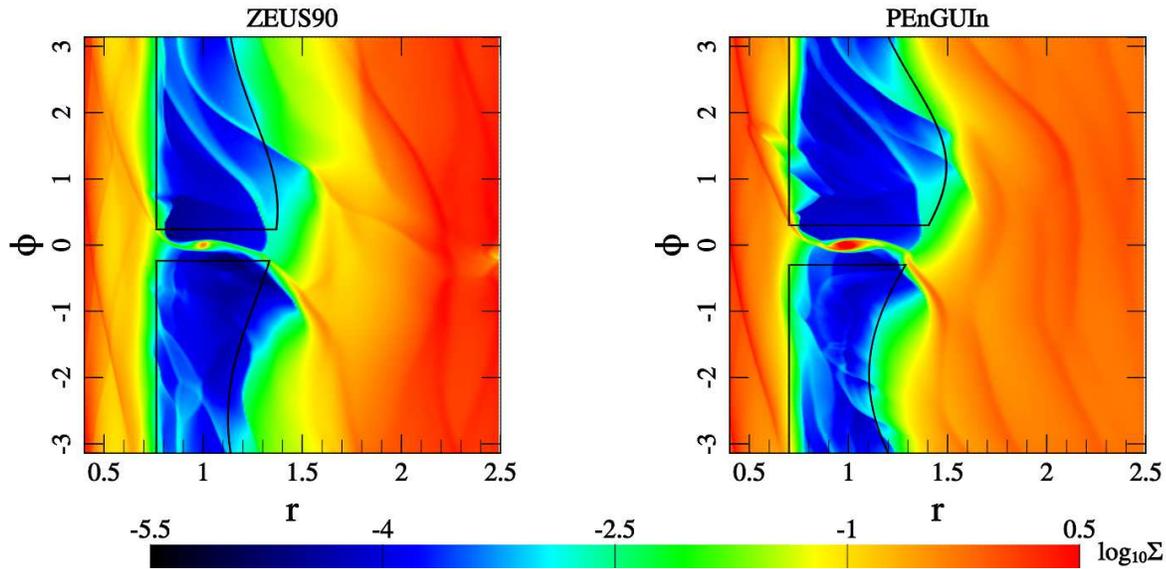}
\caption{Snapshots of eccentric outer disks, one from
  \texttt{ZEUS90} at $(q,\alpha,h/r) = (0.005, 0.001, 0.05)$,
  and another from \texttt{PEnGUIn} at $(q,\alpha,h/r) = (0.01, 0.01, 0.05)$.
  For the \texttt{ZEUS90} run shown, the inner
  edge of the outer disk (exterior to the planet's orbit)
  has eccentricity $0.10$ and precession period $630 P_{\rm p}$.
  For the \texttt{PEnGUIn} run, the eccentricity is 0.15
  and the precession period is $380 P_{\rm p}$. Black curves
  enclose the area over which $\Sigma_{\rm gap}$ is computed.}
\label{fig:eccentric}
\end{figure*}

\begin{figure*}[]
\includegraphics[width=1.99\columnwidth]{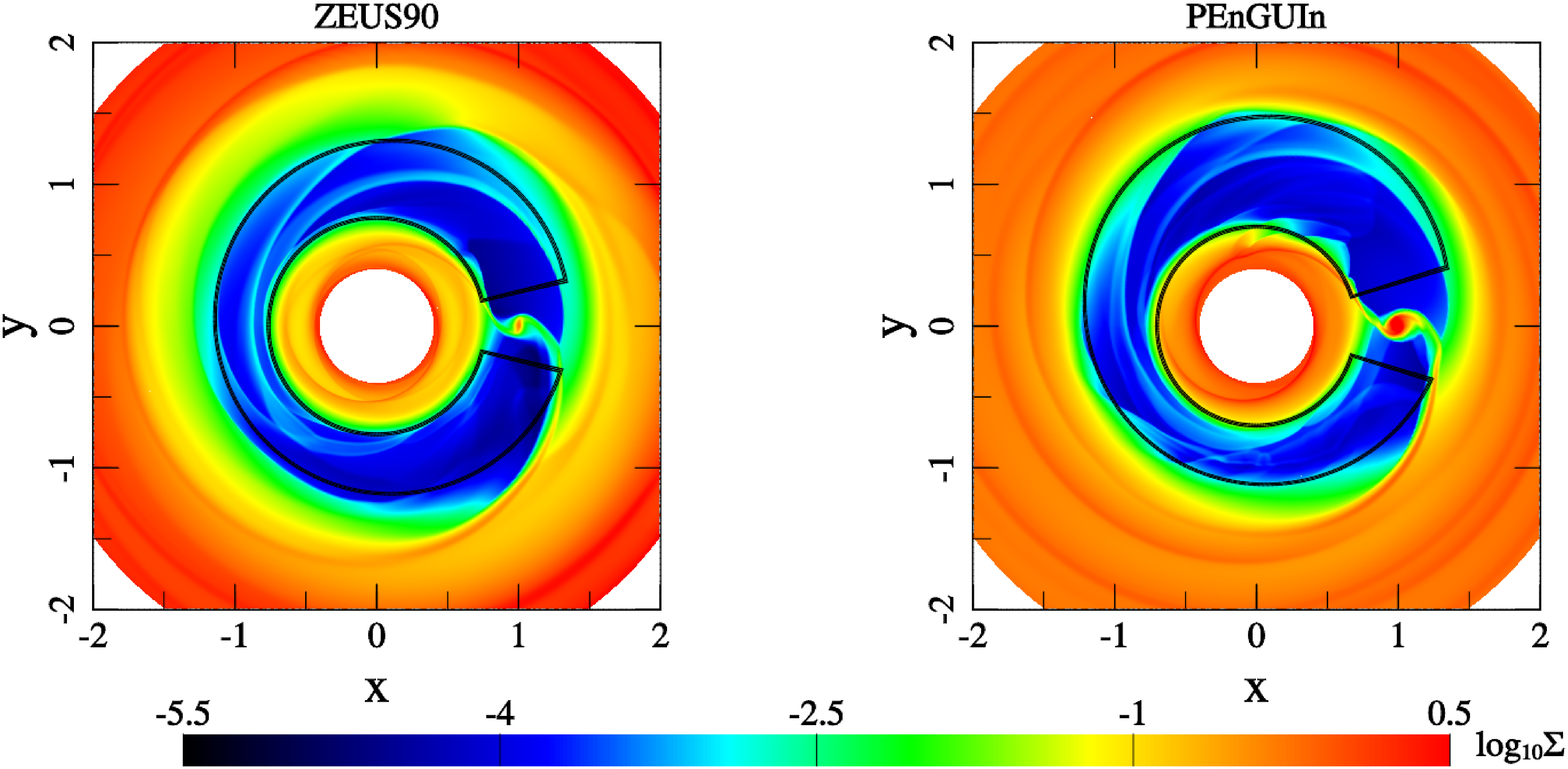}
\caption{Cartesian view of the eccentric disks of Figure \ref{fig:eccentric}. }
\label{fig:eccentric_xy}
\end{figure*}

We also observe evidence for an eccentric outer disk at high $q$;
see Figure \ref{fig:eccentric} and the ``Comments'' column in Table
\ref{tab:tab1}.  The outward transport of angular momentum by waves
launched at the 1:3 outer eccentric Lindblad resonance pumps the
eccentricity of the outer disk \citep{Lubow1991,PNM2001}.  The
$q$-value for which disks become eccentric depends on $\alpha$, $h/r$,
and disk mass \citep{Lubow1991, PNM2001, KD2006, DLB2006,
  DAA2013}. For a planet held on a fixed circular orbit embedded in a
non-gravitating disk for which $h/r = 0.05$ and $\alpha \approx
0.005$, \cite{KD2006} found that $q \gtrsim 0.003$ led to eccentric
disks; for $\alpha \approx 0.01$, the required $q \gtrsim
0.005$. Their findings are in line with ours.

In two runs with \texttt{PEnGUIn}, an
eccentricity appears in the outer disk at early times but damps away
by the time $\Sigma_{\rm gap}$ converges (see Table \ref{tab:tab1}).
The eccentricity damping is probably
an artifact of our outer circular boundary at $r_{\rm out} = 2.5 r_{\rm p}$,
which according to \cite{KD2006} is too close to the planet
to properly simulate eccentric disks. The
danger posed by the outer boundary should lessen as the
$\alpha$-viscosity increases and disturbances excited by the planet
are more localized; this may be why circularization occurs
only for our lowest $\alpha = 10^{-3}$ runs at high $q$.

We note that the fine-structure filaments threading the gaps
  are seen in most of our $q\geq0.001$ simulations, while only a few of
  these cases evince eccentric outer disks. Moreover, the filaments observed by
  \cite{devalborro06} do not appear associated with disk
  eccentricity. It is unclear to us whether the filaments and disk
  eccentricity are directly related.

\subsection{Code Comparison}\label{sec:compare}
Generally \texttt{PEnGUIn} and \texttt{ZEUS90} agree very well on
$\Sigma_{\rm gap}$ (e.g., Figures \ref{fig:q_plot}--\ref{fig:h_plot}),
typically differing by no more than a few tens of percent, and often
much better.  Because the codes rely on fundamentally different
algorithms --- one is a shock-capturing Lagrangian-remap code, while
the other is an Eulerian code using the upwind method --- their
agreement lends confidence in the accuracy of our results. Some minor,
systematic differences include: (i) $\Sigma_{\rm gap}$ for
\texttt{ZEUS90} is larger than for \texttt{PEnGUIn}; (ii)
\texttt{PEnGUIn} usually resolves a higher density peak near the
planet; (iii) \texttt{PEnGUIn} typically requires a longer time to
converge;
and (iv) near the outer boundary where
the resolution is coarser, \texttt{ZEUS90} has difficulty relaxing to
the steadily accreting solution described by equations
(\ref{eqn:init_sig})--(\ref{eqn:init_omg}), deviating from the
correct $\Sigma$ by up to 50\%.  All of these differences
may be attributed to the fact that \texttt{PEnGUIn} uses PPM, which is
a fourth-order method for uniform grids (third-order for
non-uniform grids), whereas \texttt{ZEUS90}'s algorithm is only second-order
in space.
Thus \texttt{PEnGUIn} tends to be more accurate and less
numerically diffusive than \texttt{ZEUS90}, at the cost
of taking a longer time to resolve sharp features.

A key innovation of \texttt{PEnGUIn} is its use of GPU technology to
accelerate computations.  \texttt{PEnGUIn}'s speed on a single
GTX-Titan graphics card can rival that of a traditional CPU cluster
having $\sim$100 cores. For this paper we ran \texttt{PEnGUIn} on a
desktop computer housing 3 graphics cards. With specialized hardware
we can connect up to 8 GPUs to a motherboard.  \texttt{PEnGUIn}'s
scalability with the number of cards approaches linear as the
resolution increases; our speed on 3 cards is 2.33 $\times$ that of a
single card at our standard resolution; if the resolution is doubled
(quadrupled), the speed enhancement factor increases to 2.64 (2.92) as
\texttt{PEnGUIn} takes more full advantage of GPU's
parallelism. Currently \texttt{PEnGUIn} can run on a single node only,
and would need to be modified to run on multiple nodes.  In a
multi-node cluster of GPUs, the scaling of speed with the number of
cards per node is unlikely to be linear because communication between
nodes is significantly slower than between cards on a single node.

\section{CONCLUSIONS AND OUTLOOK\label{sec:conclusion}}

We established two empirical formulas (\ref{eqn:bestfit_p}
and \ref{eqn:bestfit_highq_p}) for the surface density
contrast, $\Sigma_{\rm gap}/\Sigma_0$, inside and outside the gap
carved by a non-accreting giant planet.
The first is valid for planet-to-star mass ratios $10^{-4} \leq q \leq 5\times 10^{-3}$, and the
second is valid for $5 \times 10^{-3} \leq q \leq 10^{-2}$.  Our formulae
are derived from our new, fast, Lagrangian
shock-capturing PPM code \texttt{PEnGUIn}, and are confirmed
by \texttt{ZEUS90}.
Combining our results with those from the literature, we find that $\Sigma_{\rm gap}$
scales with $q$,
viscosity parameter $\alpha$, and disk aspect ratio $h/r$ in the following ways:
\begin{itemize}
\item At Neptune-like (and perhaps lower) masses, \cite{duffell13}
  found\footnote{A caveat is that \cite{duffell13} did not explicitly
    test the dependence on $h/r$ that they proposed. We used
      \texttt{PEnGUIn} to try to reproduce their low-mass results
      (data not shown), but encountered the problem that $\Sigma_{\rm
        gap}$ took too long to converge.  What low-mass data we did
      collect at the end of $20000$ orbital periods were consistent
      with the power-law indices proposed by \cite{duffell13}.}  that
  $\Sigma_{\rm gap} \propto q^{-2} \alpha^1 (h/r)^5$;
\item At Jupiter-like
masses, we find that $\Sigma_{\rm gap} \propto q^{-2.2} \alpha^{1.4} (h/r)^{6.6}$ (our equation \ref{eqn:bestfit_p});
\item At masses near the brown dwarf threshold, we find that $\Sigma_{\rm gap} \propto q^{-1}
\alpha^{1.3} (h/r)^{6.1}$ (our equation \ref{eqn:bestfit_highq_p}).
\end{itemize}
Our scaling indices for giant planets and quasi-brown dwarfs are
supported by two independent codes using different algorithms, and so
we are confident in their accuracy.
Note that our simulations and those of \citet{duffell13} do share one common set
of parameters: $(q,\alpha,h/r) \approx (5 \times 10^{-4}, 10^{-3},
0.05)$, for which we find $\Sigma_{\rm gap}/\Sigma_0 = 0.03$ and they
find $\Sigma_{\rm gap}/\Sigma_0 = 0.04$ (their Figure 2).\footnote{
  Technically, our simulations have spatially constant $\alpha$ and
  $h/r$, whereas theirs has spatially constant $\nu = \alpha c_{\rm s}
  h$ and $h/r \propto r^{0.25}$. Also, we compute $\Sigma_{\rm gap}$
  as an area average, whereas they report the minimum surface density.
  These differences are probably immaterial.}  We consider this good agreement.

The scaling differences between low mass and high mass,
although pointing to real physical
effects, do not lead to order-of-magnitude changes in gap depth, at
least over the range of parameters surveyed.  That is, using the
Neptune-like scaling $\Sigma_{\rm gap} \propto q^{-2} \alpha^1
(h/r)^5$ for Jupiter-like planets leads to gap depths that differ
(systematically) from those observed in our simulations by factors of
only 2--3.

\subsection{Connecting to Observations of Transition Disks}
Are the gaps empty enough to reproduce the low optical depths
characterizing the cavities of transitional and pre-transitional
disks? Surface density contrasts from models of disks like PDS 70
\citep{Dong12} and GM Aur \citep{calvet05} are $10^3$ or more. We have
found that certain sets of planet-disk parameters can achieve such
contrasts. For example, $(q,\alpha,h/r) = (5\times 10^{-3}, 10^{-3},
0.05)$ produces contrasts of $\sim$5000 (Table \ref{tab:tab1}). Lower
mass planets could also be made to work with lower
$\alpha$-viscosities and/or cooler disks with lower $h/r$; as a further example,
$(q,\alpha,h/r) = (2\times 10^{-3}, 10^{-3}, 0.04)$ generates a
contrast of $\sim$3000. The dependence on disk temperature is
especially sensitive: $\Sigma_{\rm gap} \propto (h/r)^{6.6} \propto
T^{3.3}$.

The surface density contrasts reported in this paper are
all underestimates insofar as we have neglected accretion onto the
planet; but arguably the disk accretion rate cannot be reduced by more
than a factor of order unity, lest the planet starve the host star and
violate observed stellar accretion rates (\citealt{Zhu11}; see also
\citealt{lubow06}). 
We would argue further that our gas surface density contrasts
  are also underestimates of dust surface density (i.e., optical
  depth) contrasts, to the extent that mechanisms like dust filtration
  at outer gap edges (e.g., \citealt{Zhu12}) deplete dust relative to
  gas in gaps.

Given these findings, we feel that when it comes to transition disks,
the problem is not so much gap depth, but gap width.  A single planet
embedded in an accreting disk generates a gap too narrow in radial
width to explain the expansive cavities observed in transition disks.
To connect to observations would seem to require that we expand our
study to include multiple planets or brown dwarfs within a viscous,
gravitating disk, as has been done by \citet{Zhu11}.  These authors
discounted $\alpha \lesssim 0.002$ --- and were therefore compelled to
invoke additional channels of opacity reduction (e.g., grain growth) to
explain transition disks --- because multiple planets were found to be
dynamically unstable at low $\alpha$ / high $\Sigma$ (see their page
8). The incompatibility of multiple giant planets with low $\alpha$ is a
result that we would like to see confirmed independently
and further explored.

\subsection{A Floor on $\Sigma_{\rm gap}$}
All our empirical scalings for $\Sigma_{\rm gap}$ suggest that
arbitrarily low values of $\alpha$ generate arbitrarily clean gaps. We
expect, however, the scalings to break down for small enough $\alpha$.
Without an intrinsic disk viscosity, a planet may stir the gap edge in
such a way as to trigger local instabilities and turbulent
diffusion. For example, with $\alpha=0$, the outer gap edge might be
so sharp as to be Rayleigh unstable.
There should therefore
be a minimum value or ``floor'' on $\Sigma_{\rm gap}$ caused by a
minimum planet-driven viscosity.  Just such a floor has been reported
by \cite{duffell13}; see their Figure 7.  Other simulations of
planets in inviscid disks, concentrating on orbital migration and not
gap depth, have been carried out by \cite{Li2009} and \cite{Yu2010}.

Similar arguments suggest that there might also be a floor on
$\Sigma_{\rm gap}$ at high $q$. The streamers/filaments that invade
the gap are densest for the highest $q$-values we tested.

\subsection{Analytic Derivation}\label{sec:analytic}
In \S\ref{sec:scaling}, we presented an analytic derivation of gap
depth $\Sigma_{\rm gap}$. We discovered that the power-law scalings in
our analytic relation (\ref{eqn:scaling}) match precisely those
reported from numerical experiments for low-mass planets by
\citet{duffell13}. Our analytic relation can even reproduce
approximately (deviating systematically by factors of a few) the
empirical results we found for gaps carved by Jupiter-like planets.

The success of our breezy analytic derivation for $\Sigma_{\rm gap}$
is surprising. Our derivation is ``zero-dimensional'' (``0D'') because
it considers only the total rates of angular momentum transport --
a.k.a. the total ``angular momentum luminosity'' or ``angular momentum
current'' --- integrated over all azimuth and radius; it ignores the
complicated details of how the torques are actually applied
differentially in space. We have already noted in \S\ref{sec:scaling}
how it is not completely obvious that $\Sigma_0$ characterizes
the viscous torques in gap edges.
Furthermore, our 0D treatment considers only
the wave and viscous contributions to the total angular momentum
current, and neglects the contribution from advection (i.e., the
transport of angular momentum associated with a non-zero radial
velocity $v_r$) and the contribution from azimuthal pressure variations
\citep{crida06}.

We can address some of these problems and make the leap from 0D to 1D
by considering the azimuthally averaged, radially dependent torque balance
equation for a steady-state accretion disk perturbed by a planet (see,
e.g., equation 3 of \citealt{lubow06}):
\begin{equation} \label{eqn:ld06} \frac{d(3\nu \Sigma \Omega
    r^2)}{dr} = -\Sigma \Omega r^2 v_r + 2 \Sigma r \Lambda (r)
\end{equation}
where the Lindblad torque per unit mass is
\begin{equation}
  \Lambda (r) = {\rm sgn} \, (r - r_{\rm p}) \frac{f GM_\ast q^2}{2r} \,\left( \frac{r}{r-r_{\rm p}} \right)^4 
\end{equation}
and $f$ is an order-unity constant. The left-hand side of
(\ref{eqn:ld06}) accounts for the viscous torque, while the first term
on the right-hand side accounts for angular momentum transport by
advection.  We define $x \equiv r - r_{\rm p}$ and $\dot{M} \equiv
2\pi \Sigma v_r r =$ constant $< 0$, and approximate $\Omega =
\Omega_{\rm p} (r/r_{\rm p})^{-3/2}$ so that $\nu \Omega r$ =
$\nu_{\rm p} \Omega_{\rm p} r_{\rm p} =$ constant (variables
subscripted by $p$ take their values at the planet's orbital
radius). For the outer disk, equation (\ref{eqn:ld06}) simplifies to
\begin{equation}
\alpha \left( \frac{h}{r} \right)^2 \frac{d(r\Sigma)}{dr} = 
\frac{|\dot{M}| \Omega_{\rm p} r_{\rm p}}{6 \pi GM_\ast } \left( \frac{r}{r_{\rm p}} \right)^{-1/2} 
+
\frac{f}{3} \,q^2 \, \left( \frac{r}{x} \right)^4 \, \Sigma 
\,.
\label{eqn:1D_approx}
\end{equation}
From this equation it becomes apparent how the outer accretion
disk responds when repelled outward by a planet: for $q > 0$,
the surface density gradient $d\Sigma / dr$ steepens, just enough
that the viscous torque can exceed the Lindblad torque and 
maintain a steady flow of mass inward (i.e.,
carry the $\dot{M}$ imposed at infinity across the planet's
orbit).

\begin{figure}[]
\includegraphics[width=0.99\columnwidth]{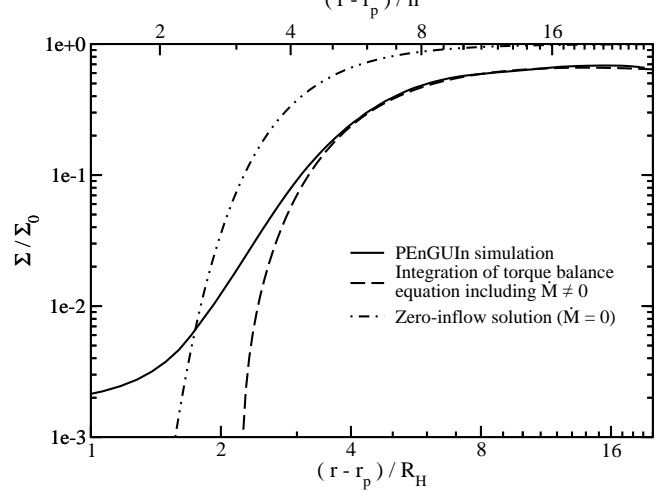}
\caption{Reproducing the simulated gap profile with a 1D analysis. The
  solid curve is the azimuthally averaged surface density profile
  outside the planet's orbit for $(q,\alpha,h/r) = (0.001, 0.001,
  0.05)$, as calculated from 2D simulations using
  \texttt{PEnGUIn}. Directly integrating the 1D equation
  (\ref{eqn:1D_approx}) reproduces well the onset of the gap, if we
  set $f=0.2$ (dashed curve). However, the bottom of the gap is not
  captured at all.  Setting $\dot{M} = 0$, as is commonly done in the
  literature, yields a profile that is similar in shape to the actual
  profile, but shifted in radius (for the same value of $f = 0.2$;
  dot-dot-dashed curve).}
\label{fig:1D}
\end{figure}

Setting $\dot{M}=0$, and working in the WKB limit where $d/dr \gg
1/r$, gives the standard zero-inflow solution: an exponential profile
for $\Sigma(r)$, commonly used in the literature (e.g.,
\citealt{lubow99}; \citealt{devalborro07}; \citealt{Mulders2013}).
Keeping $\dot{M} \neq 0$ alters $\Sigma(r)$: it still resembles an
exponential but is shifted outward (for fixed $f$), as Figure
\ref{fig:1D} demonstrates.
We find for the parameters chosen 
that equation (\ref{eqn:1D_approx}) describes well the gap profile from our
2D simulations, down to a distance of $\sim$3--4 Hill radii away
from the planet.  But inside this cut-off distance, the 1D solution
fails critically --- it falls much too steeply to recover the actual
flat-bottomed gap.  

The problem of determining the gap depth
analytically in 1D appears tantamount to the problem of understanding
what happens inside this cut-off distance.  Lindblad torques shut off
here; the tidal gravitational field of the planet is especially
strong; and circulating streamlines give way to horseshoe
orbits. One-dimensional analytic treatments may be inadequate to the
task of modeling how gas navigates from the outer disk to the inner
disk through a series of ``horseshoe turns'' \citep{lubow99,kley12}.
As far as analytic treatments go, it may be that to do better than 0D
requires at least 2D.

\acknowledgments Early explorations of this problem benefited from
Steve Lubow, Daniel Perez-Becker, and the participants of the 2011
International Summer Institute for Modeling in Astrophysics (ISIMA)
program organized by Pascale Garaud at the Kavli Institute for
Astronomy and Astrophysics in Beijing University. We thank Meredith
Hughes and Re'em Sari for encouraging discussions, and Paul Duffell,
Andrew MacFadyen, Roman Rafikov, Miguel de Val-Borro, Zhaohuan Zhu,
and especially an anonymous referee for thoughtful comments 
that led to substantive improvements. JF would like to especially
thank Pawel Artymowicz for invaluable advice. Resources supporting
this work were provided by the NASA High-End Computing (HEC) Program
through the NASA Advanced Supercomputing (NAS) Division at Ames
Research Center. Part of the simulations were performed with the
Berkeley cluster Henyey, which was made possible by a National Science
Foundation Major Research Instrumentation (NSF MRI) grant. Financial
support was provided by a NASA Origins grant.

\bibliographystyle{apj}
\bibliography{Lit}
\end{CJK*}
\end{document}